\documentstyle{article}
\input{epsf} 

\newcommand{\rem}[1]{}

\begin{document}

\title{Teleportation in a noisy environment: a quantum trajectories 
approach} 
\author{Gabriel G. Carlo, Giuliano Benenti and Giulio Casati \\ \ \\
Center for Nonlinear and Complex Systems, Universit\`a degli \\
Studi dell'Insubria and Istituto Nazionale per la Fisica della Materia,\\ 
Unit\`a di Como, Via Valleggio 11, 22100 Como, Italy}
\date{\today}

%03.67.Lx Quantum computation
%03.65.Yz Decoherence; open systems; quantum statistical methods
%03.67.Hk Quantum communication

\maketitle

%\widetext
\begin{abstract} 
We study the fidelity of quantum teleportation for the situation
in which quantum logic gates are used to provide the long distance 
entanglement required in the protocol, and where the effect of a noisy
environment is modeled by means of a generalized amplitude damping 
channel. Our results demonstrate the effectiveness of the quantum 
trajectories approach, which allows the simulation of open systems
with a large number of qubits (up to 24). This shows that the 
method is suitable for modeling quantum information protocols in 
realistic environments. 
\end{abstract}

The practical implementation of any quantum information 
protocol has to face the problem of the unavoidable 
coupling of quantum processors with their environment.
Indeed, real systems can never be perfectly isolated from 
the surrounding world. It is therefore important to understand 
the impact of the coupling with a noisy environment on the 
stability of quantum protocols. In particular, the simulation
of these protocols including realistic models of noise, 
promises to give useful insights for the design and future 
construction of quantum hardware.  

As a consequence of the unwanted environmental coupling,
a quantum processor becomes, in general, entangled with 
its environment. Therefore its state is described by 
a density matrix, whose evolution is ruled, under the 
assumption that the environment is Markovian, by a  
master equation. 
Solving this equation for a state of several qubits is  
a prohibitive task in terms of memory cost. Instead of doing 
so, quantum trajectories allow us to store only a stochastically 
evolving state vector. By averaging over many runs we get the same 
probabilities (within statistical errors) as the ones obtained 
through the density matrix directly. Therefore quantum 
trajectories are the natural approach for simulating equations 
otherwise very hard to solve.
It has been pointed out \cite{Brun} that outside the quantum 
optics and quantum foundations fields of research, the theory of 
quantum trajectories is almost unknown. 
They were used to model continuously monitored open systems 
\cite{Carmichael,DumZoller}, in numerical calculations for the 
study of dissipative processes \cite{DalibardCastin}, and in 
relation to quantum measurement theory \cite{Gisin}. 
Moreover, without using quantum 
trajectories explicitly, some studies in this spirit 
have been previously done also in quantum information \cite{MiqPaz}.
Still, there are only a few cases where they have been used 
or suggested for calculations in this area \cite{BarencoBrun,Schack}.
On the other hand, as it will become clear from the present work, 
this tool can be very powerful for the simulation of quantum 
information processing.

In this Letter, we apply the quantum trajectories formalism to 
study the fidelity of the quantum teleportation protocol 
\cite{BennettBrassard} through a large chain of qubits, in 
the presence of environmental noise. 
Quantum teleportation is a scheme by which the state of an arbitrary 
unknown qubit can be transmitted by means of a shared entangled pair,
that is an EPR pair where one particle is with the sender Alice 
and the other with the receiver Bob. Performing local operations 
and sending two bits of classical information is enough to accomplish 
the task. 
Teleportation is one of the  the basic methods of quantum communication, 
and plays a very important role in a number of quantum computation 
protocols \cite{Gottesman,KnillLaflamme}. In particular, it has been proved 
that teleportation, together with single qubit operations, 
is sufficient to construct a universal quantum computation
\cite{Gottesman}.
Several quantum optical experiments demonstrated the teleportation 
protocol \cite{Boschi,Bowmeester}, and recently a long distance 
implementation has been achieved using a 2 Km optical fiber
\cite{MarcGisin}.
Besides these rapid developments in the quantum optics arena, other  
realizations are particularly interesting from the viewpoint 
of quantum computation.
Among them, NMR experiments have been remarkably successful in 
implementing teleportation \cite{Knill}. Moreover, there are 
proposals for teleporting atomic states \cite{Bose,Polzik} and using 
quantum dot systems for electron teleportation \cite{Reina,Sauret}. 

A noiseless quantum channel is required in order for Alice 
and Bob to share a maximally entangled EPR pair, as required
by the teleportation protocol. On the other hand, the available 
quantum channels are typically noisy, as we must take into account 
the interactions of the qubits with the external world.
Recently, there has been much interest in the study of the fidelity 
of teleportation through noisy channels \cite{Oh}. 
In this work, we assume that the delivery of one of the bits 
of the EPR pair is done by means of swap gates along a noisy chain 
of qubits. As described below, this chain is an open system that 
interacts with the environment through a generalized amplitude 
damping channel. 
Using the quantum trajectories approach, we are able 
to study numerically the fidelity of teleportation, defined as the 
overlap between the final reduced density matrix of Bob's qubit 
and the original unknown state, for chain sizes of up to 24 qubits. 
We stress that the method, apart from statistical errors, is in 
principle exact, and allows the treatment of systems with a large
number of qubits, not accessible by a direct solution of the 
master equation. 
We note that a scheme for quantum teleportation in a large nuclear
spin chain has been presented in Ref.~\cite{Berman}. However, in this 
work the qubits were considered as perfectly isolated from the 
environment. 

Let us first describe the noiseless protocol. 
We consider a chain of $n$ qubits, and assume that Alice 
can access the qubits located at one end of the chain, 
Bob those at the other end. We assume that initially 
Alice owns an EPR pair (for instance we take the Bell 
state $(|00\rangle+|11\rangle)/\sqrt{2}$), while the remaining 
$n-2$ qubits are in a pure state. Thus, the 
global initial state of the chain is given by  
\begin{equation}
\sum_{i_{n-1},\ldots,i_2} 
c_{i_{n-1},\ldots,i_2} 
|i_{n-1}\ldots i_2\rangle
\otimes \frac{1}{\sqrt{2}} (|00\rangle + |11\rangle),  
\label{phiin}
\end{equation}
where $i_k=0,1$ denotes the down or up state of the qubit $k$.
In order to deliver one of the qubits of the EPR pair to Bob,
we implement a protocol consisting of $n-2$ swap gates, 
each one exchanging the states of a pair of qubits:  
$$
\sum_{i_{n-1},\ldots, i_2}
\frac{c_{i_{n-1},\ldots,i_2}}{\sqrt{2}} 
(|i_{n-1} \ldots i_2 0 0 \rangle + 
|i_{n-1} \ldots i_2 1 1 \rangle) 
$$
$$
\rightarrow
\sum_{i_{n-1},\ldots, i_2}
\frac{c_{i_{n-1},\ldots,i_2}}{\sqrt{2}} 
(|i_{n-1} \ldots 0 i_2 0 \rangle + 
|i_{n-1} \ldots 1 i_2 1 \rangle) 
\rightarrow 
$$
\begin{equation}
...\rightarrow 
\sum_{i_{n-1},\ldots, i_2}
\frac{c_{i_{n-1},\ldots,i_2}}{\sqrt{2}} 
(|0 i_{n-1} \ldots i_2 0 \rangle + 
|1 i_{n-1} \ldots i_2 1 \rangle). 
\end{equation}
After that, Alice and Bob share an EPR pair, and therefore 
an unknown state of a qubit ($|\psi\rangle=\alpha|0\rangle +
\beta|1\rangle$) can be transferred from Alice to Bob by 
means of the standard teleportation protocol 
\cite{BennettBrassard}. 
In this work, we take random coefficients 
$c_{i_{n-1},\ldots,i_2}$, that is they
have amplitudes of the order of $1/\sqrt{2^{n-2}}$ (to assure 
wave function normalization) and random phases. This ergodic 
hypothesis models the transmission of a qubit through a chaotic 
quantum channel.  

If the chain interacts with the environment, its state is 
described by a density operator $\rho$. Under the Markovian  
assumption, the dynamics of the chain is described by a 
master equation in the Lindblad form \cite{Chuangbook}: 
\begin{equation}
\dot \rho = -\frac{i}{\hbar} [H_s,\rho] - \frac{1}{2} \sum_{k} 
\{L_{k}^{\dagger} L_{k},\rho\}+\sum_{k} L_{k} \rho 
L_{k}^{\dagger},
\label{lindblad}
\end{equation}
where $H_s$ is the system's Hamiltonian,
$\{\,,\,\}$ denotes the anticommutator and  
$L_{k}$ are the Lindblad operators,
with $k \in [1,\ldots,M]$ (the number $M$ depending on the 
particular model of interaction with the environment).
The first two terms of this equation can be regarded as the evolution 
performed by an effective non-hermitian Hamiltonian, $H_{\rm eff}=
H_s+iK$, with $K=-\hbar/2 \sum_{k}L_{k}^{\dagger}L_{k}$. 
Indeed, we can see that 
\begin{equation}
 -\frac{i}{\hbar} [H_s,\rho] - \frac{1}{2} \sum_{k} 
\{L_{k}^{\dagger} L_{k},\rho\} = -\frac{i}{\hbar} 
[H_{\rm eff} \rho - \rho H_{\rm eff}^{\dagger}].
\end{equation}
The last term in (\ref{lindblad}) is usually interpreted as the one 
responsible for the so called {\it quantum jumps}. The explanation 
is as follows. If the initial density matrix describes a pure 
state ($\rho(t_0)=|\phi(t_0)\rangle \langle\phi(t_0)|$), 
then, after an infinitesimal time $d t$, it evolves into the 
following statistical mixture: 
\begin{equation}
\rho(t_0+dt)=(1-\sum_{k}dp_{k}) 
|\phi_0\rangle \langle\phi_0| + 
\sum_{k} dp_{k}  |\phi_k\rangle \langle\phi_k|, 
\end{equation}
where $dp_k = dt \langle \phi(t_0)| L_{k}^{\dagger} 
L_{k} |\phi(t_0)\rangle$, 
and the new states are defined by 
\begin{equation}
|\phi_0\rangle = 
\frac{(1-i H_{\rm eff} dt/\hbar) |\phi(t_0)\rangle}
{\sqrt{1-\sum_{k} dp_{k}}} 
\end{equation}
and
\begin{equation}
|\phi_{k}\rangle = \frac{L_{k} 
|\phi(t_0)\rangle}{||L_{k} |\phi(t_0)\rangle||}. 
\label{phimu}
\end{equation}
Then, the quantum jump picture turns out to be clear: with probability 
$dp_{k}$ a jump occurs and the system is prepared in the state 
$|\phi_{k}\rangle$. With probability $1-\sum_{k} dp_{k}$ there 
are no jumps and the system evolves according to the effective 
Hamiltonian $H_{\rm eff}$. We note that the normalization is included 
also in this case because the evolution is non-hermitian. 

To simulate numerically the evolution of the master equation 
(\ref{lindblad}), we use the so-called 
Monte Carlo wave function approach \cite{DalibardCastin},
actually implementing the above jump picture. 
We start the time evolution from a pure state $|\phi(t_0)\rangle$ and, 
at intervals $d t$ much smaller than the timescales relevant for the 
evolution of the density matrix, we choose a random number $\epsilon$ 
from a uniform distribution in the unit interval $[0,1]$.  
If $\epsilon \leq dp$, where $dp=\sum_{k} dp_{k}$, the state 
of the system jumps to one of the states $|\phi_k\rangle$
(to $|\phi_1\rangle$ if $0 \leq \epsilon \leq dp_1$,
to $|\phi_2\rangle$ if $dp_1 < \epsilon \leq dp_1+dp_2$, and so on). 
On the other hand, if $\epsilon > dp$ the evolution with 
the non-hermitian Hamiltonian $H_{\rm eff}$ takes place and 
we end up in the state $|\phi_0\rangle$. We repeat this process 
as many times as $n_{\rm steps}=\Delta t/dt$, where $\Delta t$ is 
the total evolution time. 
This procedure describes a stochastically evolving wave vector,
and we say that a single evolution is a {\it quantum trajectory}.
If we average over different runs, we recover the probabilities obtained 
using the density operator (see, e.g., Ref.~\cite{Brun}). 
Given an operator $A$, we can write the mean value 
$\langle A \rangle_t={\rm Tr} [A \rho(t)]$ as the average over 
${\cal N}$ trajectories: 
\begin{equation}
\langle A \rangle_t = \lim_{{\cal N}\to \infty}
\frac{1}{\cal N} 
\sum_{i=1}^{\cal N} \langle \phi_i(t)| A 
| \phi_i(t) \rangle. 
\end{equation}
We can see immediately the advantage of the quantum 
trajectories method: we need to store a vector of length 
$N$, where $N=2^n$ is the dimension of the 
Hilbert space, rather than a $N\times N$ matrix. 
The price to pay is that one has to run many trajectories
to get small statistical errors. 
However, a reasonably small number ${\cal N}$ of trajectories 
is sufficient to obtain a satisfactory convergence
(in our case ${\cal N}\approx 100-500$, while the maximum value
of $N$ in our simulations is $N=2^{24}$). 

We model the coupling with the environment using a 
{\it generalized amplitude damping} channel:
a state $|i_{n-1}\ldots i_0\rangle$ decays with rate $\Gamma/\hbar$. 
After an infinitesimal time $dt$, the possible states of the system 
are those in which the damping $|1\rangle \to |0\rangle$ has 
occurred in one of the qubits, the damping probability being the same 
for all the qubits \cite{damping}. 
For example, starting from the four-qubit pure state
$\rho(t_0)=|1011\rangle\langle 1011|$, the action of the generalized 
amplitude damping channel leads, after a time $dt$, to 
the statistical mixture 
$$
\rho(t_0+dt)=\left(1-\frac{\Gamma dt}{\hbar}\right)
|1011\rangle\langle 1011|+
\frac{\Gamma dt}{3\hbar}(|0011\rangle\langle 0011|
$$
\begin{equation}
+|1001\rangle\langle 1001|+|1010\rangle\langle 1010|).
\end{equation}
We would like to stress that this 
simple model must be understood as a significant example 
illustrating the power of the quantum trajectories approach
and that other kind of environmental noise can be simulated 
similarly. We assume that our quantum protocol is implemented
by a sequence of instantaneous and perfect swap gates, 
separated by a time interval $\tau$. We also assume that the 
only effect of the system's Hamiltonian $H_s$ is to generate 
these swap gates. 

Using the quantum trajectories approach, we compute numerically 
the evolution in time of the initial state vector (\ref{phiin}) 
in presence of the generalized amplitude damping channel.
The total evolution time is $\Delta t= (n-2)\tau$, since 
$n-2$ swap gates are required to transfer a member of the EPR 
pair from Alice to Bob. 
Then the standard teleportation protocol \cite{BennettBrassard}
is implemented. The fidelity of teleportation
is defined by $F=\langle\psi|\rho_B|\psi\rangle$,
where $|\psi\rangle=\alpha|0\rangle +\beta|1\rangle$ is the 
state to be teleported, and $\rho_B$ is the density matrix
of Bob's qubit at the end of the teleportation protocol, 
obtained from the final state of the $n$-qubit chain after 
tracing over all the other qubits of the chain. 
In the quantum trajectories method, we compute the fidelity as 
\begin{equation}
F = \lim_{{\cal N}\to \infty}
\frac{1}{\cal N} 
\sum_{i=1}^{\cal N} \langle \psi | (\rho_B)_i 
| \psi \rangle,
\end{equation}
where $(\rho_B)_i$ is the reduced density matrix of 
Bob's qubit, obtained from the final $n$-qubit state of 
the trajectory $i$. 

\begin{figure}[ht]
\centerline{\epsfxsize=8cm\epsffile{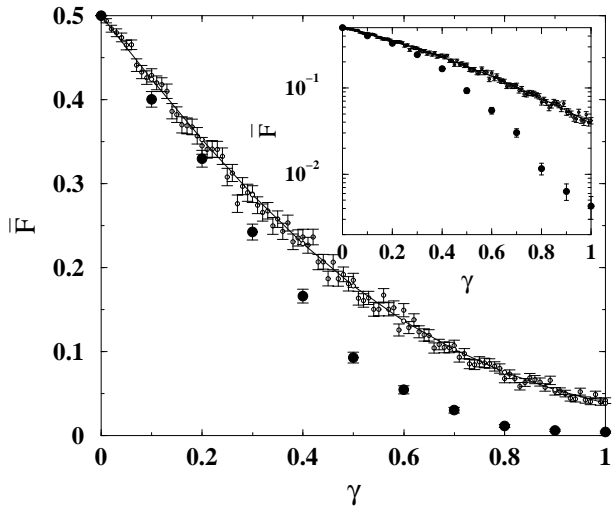}}
\caption[]{\footnotesize Fidelity ${\bar F}=F-F_\infty$ of the 
teleportation of the state $|\psi\rangle = (|0\rangle + |1\rangle)/\sqrt{2}$
as a function of the dimensionless damping rate $\gamma$, for
$n=9$ (circles) and $n=22$ qubits (filled circles). 
The curve is obtained by direct solution of the master
equation at $n=9$, while the circles give the results of the
quantum trajectories approach with ${\cal N}=400$ trajectories.
Inset: the same but with a logarithmic scale for $\bar{F}$.
The error bars give the size of the statistical
errors.}
\label{fig1}
\end{figure}

\begin{figure}[ht]
\centerline{\epsfxsize=8cm\epsffile{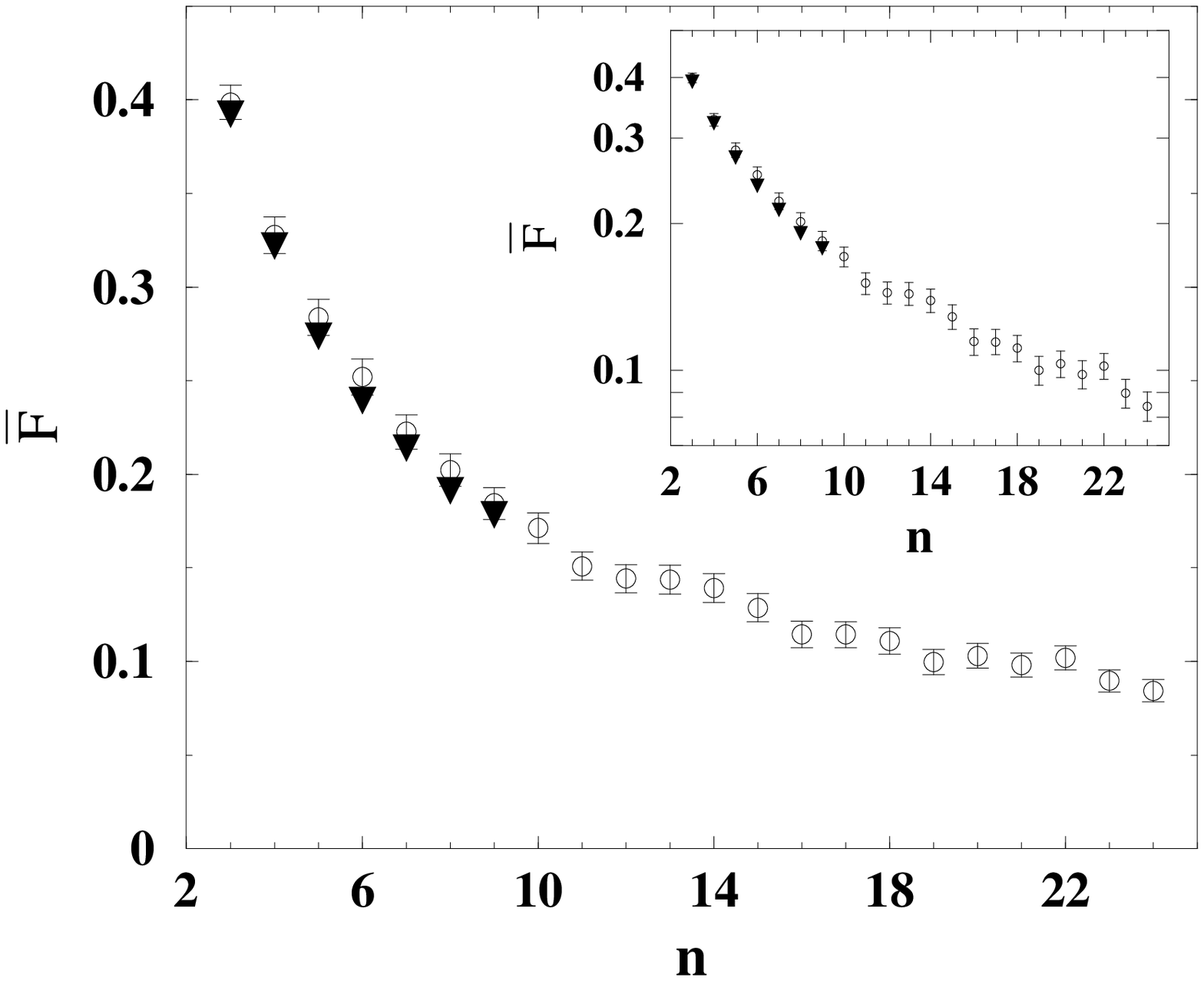}}
\caption[]{\footnotesize Fidelity as a function of 
the number of qubits in the chain, for $\gamma=0.5$. 
The state to be teleported is 
$|\psi\rangle = (|0\rangle + |1\rangle)/\sqrt{2}$.
The results are obtained from the quantum trajectories 
method with ${\cal N}=400$ (circles) and from direct 
integration of the master equation (triangles). 
Inset: the same but with a logarithmic scale for $\bar{F}$.}
\label{fig2}
\end{figure}

The comparison between the quantum trajectories approach and 
the direct solution of the master equation (\ref{lindblad}) 
is shown in Fig.~\ref{fig1}, where we compute the 
fidelity of the teleportation protocol as a function of 
the dimensionless damping rate $\gamma=\Gamma\tau/\hbar$.
As an example, we consider the teleportation of the 
state $|\psi\rangle=(|0\rangle+|1\rangle)/\sqrt{2}$.
Since the amplitude damping channel is such that the 
system ends up, at times much longer that $\hbar/\Gamma$, 
in the state $|0\ldots 0\rangle$, we expect that for 
this state the fidelity drops to the value $F_\infty=1/2$. 
This is confirmed by the numerical data of Fig.~\ref{fig1},
where we can see that the quantum trajectories method correctly 
reproduces the decay of $\bar{F}\equiv F-F_\infty$.
We point out that already with a rather small number of 
trajectories, ${\cal N}=400$, statistical errors 
are sufficiently small to capture the relevant physical 
features of the model. It can be seen from Fig.~\ref{fig1}
that the agreement with the results obtained from the direct 
integration of the master equation (\ref{lindblad}) is remarkably 
good. The most important point is that the quantum trajectories
approach allows us to simulate a number of qubits much 
larger than those accessible by direct solution of the 
master equation, which, due to memory restrictions in a 
classical computer, is possible only up to $n\approx 10$ qubits.
The decay of the fidelity with the length $n$ of the qubit
chain is shown in Fig.~\ref{fig2}. Again we note that 
the agreement between quantum trajectories and direct 
numerical solution of the master equation is satisfactory.
However, with the first approach we are able to simulate
much longer spin chains with up to 24 qubits. 
We note that, as we will discuss in detail elsewhere, 
the non exponential decay of the fidelity with the damping rate 
and the number of qubits is a feature of the generalized amplitude 
damping channel.
We would like to stress that this non trivial decay is 
correctly reproduced by the quantum trajectories approach. 

In summary, we have studied the fidelity of the teleportation
protocol in a noisy environment by means of the quantum trajectories
method. Our studies demonstrate the ability of this approach 
to model quantum information protocols with a large number of 
qubits. This opens up many possibilities for
future studies. Theoretical predictions for the behavior 
of the fidelity and of other relevant quantities 
with respect to the system size and different kinds 
of environment can now be explored with the help of numerical
simulations. It will be also possible to include the effects of
realistic internal Hamiltonians. Since various quantum 
protocols can be easily modeled, the scope
of the present approach can be extended to the study of their
stability.
Finally, quantum trajectories offer a very convenient
framework to model experiments. In this context, we point out 
the ability of a single quantum trajectory to provide
a good illustration of an individual experimental run
\cite{DalibardCastin}.
Therefore quantum trajectories promise to become a very 
valuable tool for quantum hardware design and to 
determine optimal regimes for the operability of 
quantum processors. 

%%\begin{acknowledgments}

This work was supported in part by the EC contracts 
IST-FET EDIQIP and RTN QTRANS, the NSA and ARDA under
ARO contract No. DAAD19-02-1-0086, and the PRIN 2002 
``Fault tolerance, control and stability in
quantum information precessing''.

%%\end{acknowledgments}

\bibliographystyle{prsty}
%\bibliography{}
%\bibliography{../bib/fg4,../bib/extern}

\begin{thebibliography}{99}
\bibitem{damping} It is possible to work out a compact expression for 
the Lindblad operators $L_k$ in our model. Their matrix elements
in the basis $|i\rangle \equiv |i_{n-1}\ldots i_0\rangle$, with 
$i\equiv\sum_{l=0}^{n-1} i_l 2^l$, are given by   
\begin{equation}
[L_k]_{i,j} = \left\{
\begin{array}{ll}
 \sqrt{\frac{\Gamma}{\hbar \sum_{l=0}^{n-1} i_l}}, & 
\;{\rm for}\; j\geq 2^k, \; i=j-2^k, \; i_l=1, \\ 
 & \\
 0  & \;{\rm otherwise}. 
\end{array} \right.
\end{equation}
There are $n$ operators $L_k$ ($k=0,1,\ldots,n-1$), 
where the index $k$ singles out
which qubit undergoes the transition $|1\rangle \to |0\rangle$.
\end{thebibliography}

\end{document}